\documentstyle[12pt]{article}
\input{psbox.tex}
\title{A possible solution of the black hole information paradox\\
through quantum gravity unified with other interactions}
\author{Horatiu Stefan Nastase\\Niels Bohr Institute\\Blegdamsvej 17, 2100 
Copenhagen \O \\Denmark}

\date{January 9, 1996}
\begin{document}
\maketitle
\begin{abstract}
I try to argue that the only way out of the black hole information paradox
is through a unified quantum field theory of gravity and other interactions.
Superstring theory is especially interesting, since in a special limit, the
classical picture of 't Hooft emerges. 
\end{abstract}
\section{Introduction}
One of the most puzzling problems in high energy physics is the black hole 
information paradox, which can be stated simply as the fact that the collapse 
of a pure quantum state of matter forming a black hole leads to a thermal 
(mixed) state of radiation, completely uncorrelated and independent of the 
initial state, and this conclusion is apparently based only on low energy 
physics, without having to say anything about high energy physics. 

Moreover, apparently every possible outcome of the black hole evaporation:
pure state, remnant (stable, Planck scale object, which doesn't further decay,
keeping the information about the initial state) and also the thermal state,
lead to violation of laws we think should apply. That's why we say we have a 
paradox (no solution consistent with low energy physics) and not merely a
problem.

I will try to argue that this is not the case, and if we have a pure initial
state, the final state could also be pure, if nonperturbative quantum gravity
effects are such as to preserve unitarity. Moreover, if we assume that the 
initial information of the quantum state should be given by spins, momenta
and charges, the condition that part of the information be regained before
the end of the black hole evaporation leads to the necessity of having a 
unified theory of quantum gravity and other interactions at the Planck scale
(nonunified quantum gravity doesn't regain the information).

A perturbative superstring computation is shown to satisfy the requirements.

In section 2 I give a brief review of the arguments why we have a paradox.
In section 3 I argue about the points where we can find flaws in the usual
arguments: particles get created at the collapsing star's surface, in some
sense, due to nontrivial transformation laws for coordinates, their creation
being {\em associated} with the horizon, but not in a local way -suggested by
the Hawking formula (temperature proportional to surface gravity)- and at their
birth, particles come with a huge momentum. Moreover, we have an infinite boost
between infalling and Schwarzschild frames, implying a black hole 
complementarity. In section 4 I present the general model: 't Hooft's classical
picture of the scattering of initial state into Hawking radiation is replaced
by a picture of unified quantum field theory of gravity and other interactions,
where Hawking particles never leave the star's surface in a mixed state, but 
instead interact (in a way essentially nonperturbative) with the star, getting
correlated with the incoming particles and the states inside the horizon. A 
useful approximation is an S matrix, allowing us to see the general features. 
Unlike the Unruh effect, energy density and energy flow is consistent with 
the existence of particles in both reference frames, and the complementarity 
of the picture is due to the combination of gravitational interaction and
infinite boost. In section 5 I present a superstring computation -along the
lines of a research by Amati et al.- which suggests that the classical 
picture of 't Hooft is a special limit of the superstring case, whereas I point 
out that the superstring case information can be regained before the endpoint.

\section{The information paradox}

Hawking showed (\cite{haw1,haw2})
 that black holes radiate a thermal radiation (mixed state),
completely independent on the initial state. However, even in its original
formulation, the absence of an appropriate way to introduce back-reaction on
the metric -the loss of energy of the black hole through radiation implies
that the black hole should shrink in size, and this effect has to be taken into 
account in a proper way- left the possibility to have -maybe- a regain of this
initial information.

\begin{figure}
$$\psbox{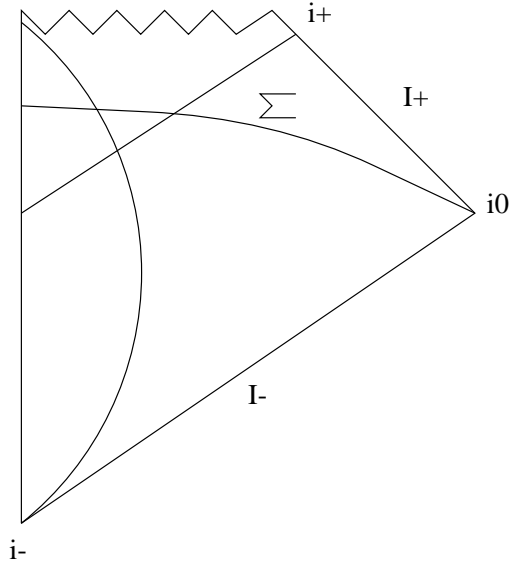}$$
\caption{spacelike slice cutting through the inside and outside of the 
black hole}
\label{fig1}
\end{figure}

This could apparently not happen before the black hole becomes planckian, for
basicly 2 reasons:

-on a spacelike $\Sigma$ (see figure \ref{fig1}) 
which cuts through the inside and the outside of the
black hole, we should have independence of the observables inside and outside 
the black hole, explicitly if $x\in inside$, $y\in outside$,
\begin{equation}
[ O(x),O(y)]=0
\end{equation}
in the case of a local and causal theory, or at least falling off exponentially:
\begin{equation}
[ O(x),O(y)]\sim e^{-\frac{(x-y)^{2}}{l_{Pl}^{2}}}
\end{equation}
so that it will not matter, when quantum gravity effects are taken into account.

-the "no quantum Xerox" principle, stating that we can't copy a quantum state,
because this violates the superposition principle, which is a basic requirement
of quantum theory:
\begin{eqnarray}
&&|A>\rightarrow |A>_{inside}\times |A>_{outside}\Rightarrow\nonumber\\
&&|A>+|B>\rightarrow |A>\times |A> +|B>\times |B>\neq (|A>+|B>)\times (|A>+|B>)
\end{eqnarray}

However, both arguments are in some sense unessential, since we have in 
ordinary quantum mechanics, the EPR paradox, whose effect is more or less
the same: in its simplest form, it says for instance that when you have an
interaction describing the decay of a spinless particle at rest into 2 particles
with spins, the spins must always be opposite, i.e. $|s=0>\stackrel{int}{
 \rightarrow}|\uparrow>|\downarrow>+|\downarrow>|\uparrow>$, meaning that the
final particles are correlated, and measuring the spin of one particle is 
equivalent to measuring the spin of the other, even if the other is at a huge 
distance (the price is, however, the collapse of the wavefunction). I will
try to argue about a similar thing in what follows.

Anyway, if information about the initial state would not be available before 
the planckian stage, we seem to have again a trouble:

-if information doesn't return, as Hawking suggests, there appear to be 2
types of problems:

a)we should either drop the quantum mechanical superposition principle, when 
ther is the possibility of black hole formation, or drop the clustering property
of probabilities of black hole formation at different points (
\cite{stro,hastro}): If we consider 
the state $|\psi>=1/\sqrt{2}(|x_{1}>+|x_{2}>)$, where $|x_{1}>$ forms a black
hole at $x_{1}$ and $|x_{2}>$ at $x_{2}$, in the resulting superscattering 
operator \$  mapping density matrices: $\rho_{out}=\$ \rho_{in}$, we should 
consider the "diagrams" corresponding to $|x_{1}><x_{1}|,|x_{2}><x_{2}|$, but 
also the ones corresponding to $|x_{1}><x_{2}|$ and $|x_{2}><x_{1}|$, from the 
superposition principle, but then for the state $|x_{1}>|x_{2}>$ we should add 
to $|x_{1}><x_{1}||x_{2}><x_{2}|$ also $|x_{1}><x_{2}||x_{2}><x_{1}|$, meaning 
that I must have 
correlation between the forming of black holes at widely separated locations,
which is the opposite of clustering ($P_{x_{1}x_{2}}=P_{x_{1}}P_{x_{2}}$).
So knowledge of all past and future black hole formation is required to 
compute the superscattering matrix for a single experiment.

b)the second problem is that then information loss could take place in virtual
processes, which would mean that the world is in contact with a heat bath at 
$T\sim T_{Pl}$.

-if we consider the existence of remnants, either stable or long lived, so as
to radiate all information about the initial state, the problem which arises
is that apparently you should have an infinite number of species of remnants,
corresponding to the possible types of infalling matter.(\cite{stro,hastro,gid})
 Then, since remnants 
have mass, it should be possible to pair create them in a gravitational
field and, naively, the pair production rate should be proportional to the 
number of species, therefore infinite.

\section{Suggestions}

Let's look now how the Hawking radiation appears, what are the conditions to 
put in order to get Hawking radiation, and what are the points where 
something can go wrong.

Apparently, the Hawking radiation is a charactersitic of a very general type 
of gravity models, and I am not talking only about models which give rise 
to a singularity (the types of models to which the Hawking and Penrose theorems
(\cite{hp1,hp2,hp3,hp4,hp5}) apply), 
but about models which give rise to a horizon, and where the matter
collapses through this horizon.

As is well known, Hawking radiation arises because of the existence of a 
Bogoliubov transformation between the incoming and outgoing modes:

For a scalar field, if $\{f_{i}\}$ is a complete orthonormal basis of the 
solution of the field equation ($\phi_{;ab} g^{ab}=0$, e.g.) on the past null
infinity $I^{-}$ (incoming modes), and $\{p_{i}\}$ solutions of the field
equation purely outgoing, and $\{q_{i}\}$ solutions with no outgoing component,
then: 
\begin{eqnarray}
\phi=\sum_{i}\{f_{i}a_{i}+\bar{f_{i}}a_{i}^{+}\}&=&\sum_{i}\{p_{i}b_{i}+\bar{p_
{i}}b_{i}^{+}+q_{i}c_{i}+\bar{q_{i}}c_{i}^{+}\}\\
p_{i}=\sum_{j}(\alpha_{ij}f_{j}+\beta_{ij}\bar{f_{j}})&;&q_{i}=\sum_{j}(\gamma
_{ij}f_{j}+\eta_{ij}\bar{f_{j}})\;and\\
b_{i}=\sum_{i}(\bar{\alpha_{ij}}a_{j}-\bar{\beta_{ij}}a_{j}^{+})&;& c_{i}=
\sum_{j}(\bar{\gamma_{ij}}a_{j}+\bar\eta_{ij}a_{j}^{+})
\end{eqnarray}
and as we can see the existence of the $\beta$ coefficient implies particle
production, i.e. $|0>_{in}$, defined by $a_{in}|0>_{in}=0$ differs from the out 
vacuum $|0>_{out}$:
\begin{equation}
_{in}<0|b_{i}^{+}b_{i}|0>_{in}=\sum_{j}|\beta_{ij}|^{2}
\end{equation}
That means that if we find that $p_{i}$ has a component corresponding to 
$\bar{f_{j}}$, we will have particle creation. In order to see that, we 
propagate backwards the solution $p_{i}$ at future infinity, which is 
proportional to the factor $e^{-i\omega u}$. Following Birrell\&Davies (\cite
{bd}), we
write the general case of a metric, outside the collapsing ball of matter,
\begin{eqnarray}
ds^{2}&=&C(r)dudv\\
u&=&t-r^{*}+R_{0}^{*}\\v&=&t+r^{*}-R_{0}^{*}\\r^{*}&=&\int C^{-1}dr
\end{eqnarray}
and inside the ball:
\begin{eqnarray}
ds^{2}&=&A(U,V)dUdV\\U&=&\tau-r +R_{o}\\V&=&\tau+r-R_{0}
\end{eqnarray}
and $C\rightarrow 1, \frac{\partial C}{\partial r}\rightarrow 0$ as $r
\rightarrow \infty$. We assume also that there is an event horizon in the 
outside metric, where C=0, and that the ball collapses along $r=R(\tau)$ 
with $r(\tau=0)=R_{0}$.

\begin{figure}
$$\psbox{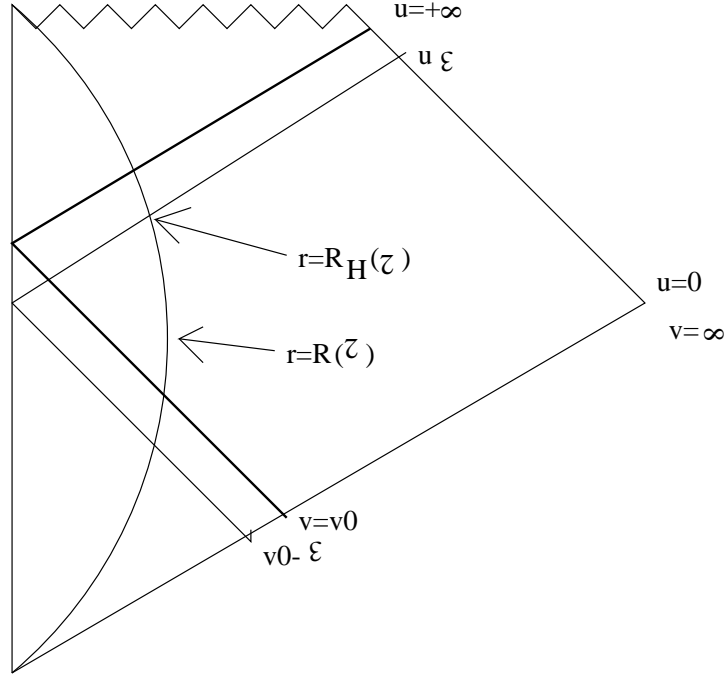}$$
\caption{Penrose diagram of a collapsing black hole}
\label{fig2}
\end{figure}

If we have transformation equations between the interior and exterior 
coordinates, $U=\alpha(u)$ and $v=\beta{V}$, then the matching conditions
across $r=R(\tau)$ , 
\begin{equation}
ds^{2}_{r=R(\tau)}=C(r)dudv=A(U,V)dUdV
\end{equation}
imply, after some computations,
\begin{eqnarray}
\alpha'(u)|_{r=R(\tau)}&=&\frac{dU}{du}|_{r=R(\tau)}=(1-\dot{R})C\{[AC(1-\dot{
R}^{2})+\dot{R}^{2}]^{1/2}-\dot{R}\}^{-1}\\ \beta'(V)|_{r=R(\tau)}&=&\frac{dv}
{dV}=C^{-1}(1+\dot{R})^{-1}\{[AC(1-\dot{R}^{2})+\dot{R}^{2}]^{1/2}+\dot{R}\}
\end{eqnarray}
Looking at the Penrose diagram of the collapsing black hole (fig.\ref{fig2}) 
, we see that a mode
$e^{-i\omega u}$ (at u=ct.), propagated backwards in time, passing through the
collapsing body, reflects on the r=0 origin of coordinates and gets out near
$v=v_{0}$, the last incoming line which can escape to future null infinity
without being trapped inside the black hole. So, a range $u=u_{0}$ up to 
$u=\infty$ corresponds to a range $v_{0}-\epsilon$ up to $v_{0}$, so almost
all the outgoing radiation corresponds to a small region near $v_{0}$.

Thus, we can make 2  simplifications in our case, to solve the matching
constraints:

We are interested in $U=\alpha(u)$ matching condition for the mode $e^{i\omega u
}$ in a small region of $r=R(\tau)$ near the event horizon C=0, corresponding
to late retarded times ($u\rightarrow \infty$), so relation simplifies
near C=0 to
\begin{equation}
\frac{dU}{du}|_{r=R(\tau)}\simeq -[U+R_{h}-R_{0}-\tau_{h}]k
\end{equation}
where $k\stackrel{def}{=}1/2\frac{\partial C}{\partial r}|_{r=R_{h}}$ is the 
surface gravity of the black hole, and we put $R(\tau)\simeq R_{h}+\nu(\tau-
\tau_{h})$, $\nu=-\dot{R}(\tau_{h})$ and $R_{h}(\tau_{h})$ correspons
to horizon crossing.

Integrating this equation, we get 
\begin{equation}
U\simeq E\;e^{-ku}+F
\end{equation}
, on $r=R(\tau)$ and near the horizon (C=0).

Similarly, we are interested in the $v=\beta(V)$ matching condition for the 
small region of $r=R(\tau)$ near $v=v_{0}$, and if we further approximate 
that this is also near the horizon (C=0), so 
\begin{equation}
\frac{dv}{dV}\simeq\frac{A(1-\dot{R})}{2\dot{R}}
\end{equation}
and since v varies very few, A$\simeq$ const. and $\dot{R}\simeq \dot{R}(\tau_{
h})=-\nu$, so that
\begin{equation}
v\simeq const.-\frac{AV(1+\nu)}{2\nu}
\end{equation}
Now let us see what is going on: we have the outgoing modes (basis vectors),
proportional to $e^{i\omega u}$. This form of the mode is constant outside the
collapsing ball. When we arrive at the surface of the ball we must use the 
transformation $U=\alpha(u)$. At the $r=0$ origin of coordinates we have $V=
U-2R_{0}$, and then when we get out to past infinity, we must use the form of 
$v=\beta(V)$. So the relation between the coordinates labelling the modes 
at past and future null infinity is:
\begin{equation}
v=\beta(V)=\beta(U-2R_{0})=\beta(\alpha(u)-2R_{0})
\end{equation}
So if we have a mode that is incoming at $I^{-}$, it will have the form
\begin{equation}
\frac{1}{\sqrt{4\pi\omega}}(e^{-i\omega v}-e^{-i\omega\beta[\alpha(u)-2R_{0}]})
\end{equation}
or, in our case (with the necessary approximations):
\begin{equation}
\simeq\frac{1}{\sqrt{4\pi\omega}}(e^{-i\omega v}-e^{i\omega(E'e^{-ku}+F')})
\end{equation}
The outgoing mode is found by inverting $p(u)=E'e^{-ku}+F'$ to $p^{-1}(v)=
f(v)=-\frac{1}{k}\log \frac{v-v_{0}}{C}$, so the outgoing mode is
\begin{equation}
\frac{i}{\sqrt{4\pi\omega}}(e^{i\frac{\omega}{k}\log \frac{v-v_{0}}{C}}-e^{
-i\omega u})\;\;(v<v_{0})\label{phase}
\end{equation}
Up to now, nothing new (the computation is in Birrell and Davies (\cite{bd})). 
However,
we must stop for a moment and look at the assumptions made and physical
implications: The only assumption made is that the outside metric has a 
nontrivial solution to the equation C=0, corresponding to an event horizon
and that the matter ball collapses through this event horizon. In order to
derive the exact form of the outgoing modes, we had to use some approximations,
but this is unessential.

Another, more recent, type of computation, relates Hawking radiation also to
the moment of collapsing trough the event horizon, but from the point of vue
of what a detector feels (\cite{fh,hb})  

Had we chosen that the matter remains at $r=R(\tau)=ct.=R$, we would have
obtained:
\begin{equation}
\frac{dU}{du}|_{r=R}=\sqrt{\frac{C(r)}{A(U,V)}}\;and\;\frac{dv}{dV}|_{r=R}=
\sqrt{\frac{A(U,V)}{C(r)}}
\end{equation}
and if A(U,V)=A(r) (seems logic), then A(U,V)=const., C(r)=const., so U=Cu+D,
v=$\frac{1}{C}$V+E, and we would have a linear relation v=u+const., implying
no $\beta$ coefficient ($e^{i\omega u}$ becomes $e^{i\omega v}$, up to 
normalization), so no particle creation. The nontrivial phase in eq.\ref{phase},
 allowing a 
$\beta$ coefficient, and leading, as Hawking showed, to a thermal state, with
$T=\frac{k}{2\pi}$, comes because we have an event horizon (C=0) and the matter
collapses through it: a small zone in v, $[v_{0}-\epsilon,v_{0}]$ is mapped
to a huge zone in u: $[u_{\epsilon}, \infty]$, by an exponential depending
on the surface gravity of the hole. 

One thing we neglected until now (the most important in my opinion) is the fact 
that although the mode {\em form} is the same, the frequency $\omega$ is not.
When we propagate backwards from future null infinity, $\omega$ is blueshifted
with a factor increasing with the retarded time u, so that at $u\rightarrow
\infty$, the mode gets close to the horizon (C=0) and is usually (since
Hawking) suggested that {\em because of this} ($\omega|_{r=R(\tau)}\rightarrow
\infty, \lambda\rightarrow 0$), we can use the geometric optics approximation
for the {\em wave}, and propagate on a straight line, with no distorsion of 
the mode form either, through the collapsing body and out to $I^{-}$, when we
have a corresponding redshift, so that $\omega|_{I^{-}}=\omega|_{I^{+}}$.

Before I comment on it though, I will say that I think we have here a quite 
different thing from the Unruh radiation, although it is often connected with
it. Unruh radiation appears when we have an accelerated observer in Minkowski
space, so that although for the Minkowski observer we have a vacuum state
$|0>_{M}$, for the accelerated one we have a thermal state, "$|th.>_{acc.}$",
different from the vacuum in his reference frame. But then both observers
agree that $_{M}<0|:T_{\mu\nu}:|0>_{M}=0$ (respectively, $_{M}<0|:T_{\mu\nu}':|
0>_{M}=0$) and the accelerated detector feels thermal radiation because the 
acceleration induces a transition in both the field and the detector (Birrell\&
Davies). The thermality arises there because of the ambiguity of the particle
concept in a field theory in curved spacetime (its global nature-a particle
is defined on all spacetime). In a general spacetime, there is no preffered
reference frame where to define physical particles.

In the black hole case, first of all, we have real flux of particles at $I^{-}$,
so we have an energy density and energy flux, besides particle number (which 
could have been considered irrelevant). Secondly, if we propagate backwards
the whole thermal flux, because of the blueshift, the temperature increases 
by $\sqrt{C(r)}$: $T=\frac{k}{2\pi}\frac{1}{\sqrt{C(r)}}$, so that near the
horizon we have $T=\frac{1}{2\pi}\frac{1}{2}\frac{\partial C}{\partial r}|_{
r=R_{h}}\frac{1}{\sqrt{C(r)}}$, and we see that  this corresponds to the Unruh 
temperature $T=\frac{a}{2\pi}$, because the gravitational acceleration $\vec{g}$
is defined as $\vec{g}=-\vec{\nabla}\log \sqrt{C(r)}$, where $\vec{\nabla}=
\vec{e}_{r}\sqrt{C(r)}\frac{\partial}{\partial r}+angular$, so that 
$\vec{g}=-\frac{1}{2}\frac{1}{\sqrt{C}}\vec{e}_{r}\frac{\partial}{\partial r}
C(r)$. This implies that near the horizon $T\simeq\frac{|\vec{g}|}{2\pi}$, 
which would
suggest us to say that the freely falling observer is the physical one, seing 
no radiation, so the fiducial observer (Schwarzschild) sees a Unruh radiation
$T\simeq\frac{|\vec{g}|}{2\pi}$, which gets out redshifted by $\sqrt{C(r)}$, 
so $T=\frac{\sqrt{C(r)}|\vec{g}|}{2\pi}|_{horizon}=\frac{k}{2\pi}$. 
The fact that 
the reasoning is not quite true can be seen quite easily because then we could 
say equally well that for a star which doesn't collapse we have $T=\frac{k}{2
\pi}$, where k is defined at the surface of the star. We have seen that this
is not true, and the explanation can be seen as being the {\em globality}
of the particle concept, not the locality apparent in k (defined at the horizon)
, i.e. the real cause of particle production is traced as being inside the star,
where we put nontrivial transformation laws for the coordinates inside/outside.
The nontrivial transformation laws are {\em associated} with the horizon, but it
is inside the star that particles get created, in some sense. (note that I am 
not coming back to locality, because the appearence of the particles is 
connected to both the collapsing star and the horizon, so it is global, just 
that outside the star we have particles, whereas inside we cannot really say)

\begin{figure}
$$\psbox{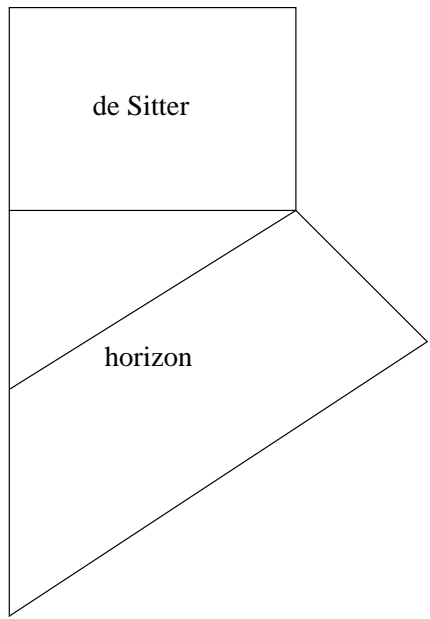}$$
\caption{black hole in a model with a minimum curvature- inside the black hole, 
the matter reexpands through a de Sitter phase}
\label{fig3}
\end{figure}

One more observation is needed: since the particle creation is seen to be 
correlated to the existence of the horizon, and has nothing to do with the
region where the curvature becomes planckian and full quantum gravity should
apply, we are led to say that apparently Hawking radiation appears independently
of the details of Planck scale physics: as we saw, the only requisite was the
existence of the horizon (C=0 in above). In particular, for a model such as 
the one studied by Brandenberger (\cite{br}), where there is a 
limiting curvature around the Planck 
scale, and inside the black hole the matter collapses and then reexpands 
(see fig.\ref{fig3})-by
a de Sitter phase- (that is how it manages to avoid Hawking and Penrose
singularity theorems), the existence of the horizon is enough to ensure the
same argument to hold.

But now the problem becomes more apparent: The assumption which was made was
that we can think of modes as waves, and that we have no refraction or
dispersion, so that the only effect to be taken into account is the 
transformation $\{ U=\alpha(u),v=\beta(V)\}$ across the star's surface. However,
the transformation of the modes $p_{i}=\sum_{j}(\alpha_{ij}f_{j}+\beta_{ij}
\bar{f}_{j})$ is associated with a transformation of the creation operators
$b_{i}=\sum_{j}(\bar{\alpha}_{ij}a_{j}-\bar{\beta}_{ij}a_{j}^{+})$ so that we
should not think of the particular modes which we propagate backwards only
as waves, but also as particles, and then the blueshift implies that these 
particles get boosted to a huge momentum, and if we consider a particle as
having some uncertainty in momentum at $I^{+}$ so as to have a localization 
around a particular wavefront u, the boost increases with u. That means that 
this particle gets out of the star (since we argued that particles get created
inside the star) with a huge momentum= boost relative to the collapsing matter.
In fact, for the cases of interest, the boost gets well above the Planck 
scale.

Then this picture of propagation by geometric optics seems to me that looses
its power, because we have found a point where superplanckian  physics should
enter the game.

\begin{figure}
$$\psbox{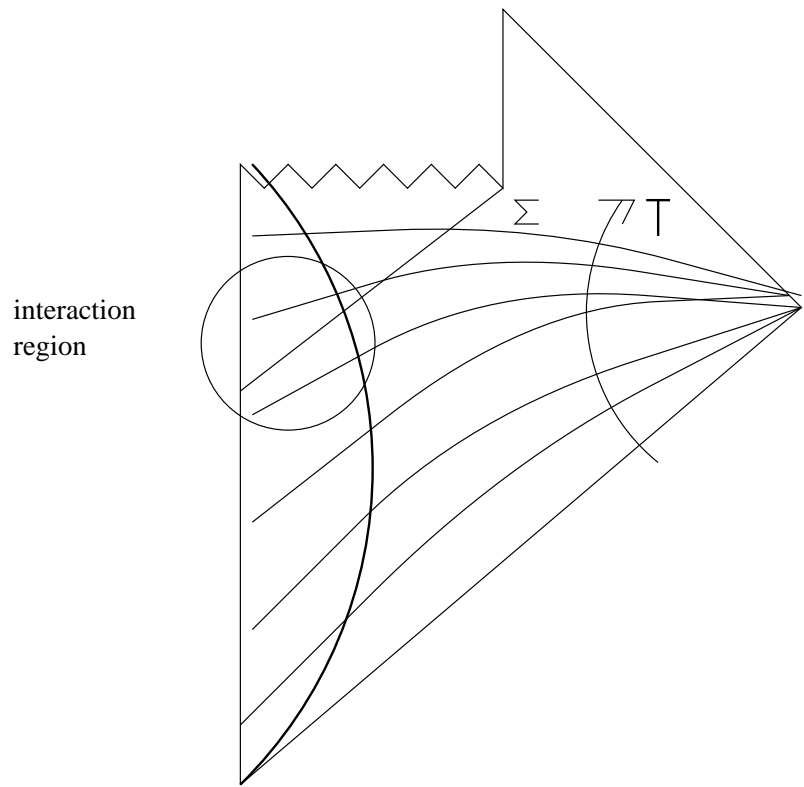}$$
\caption{the "adiabatic time slicing", where the adiabatic slices are 
labelled by a time T}
\label{fig4}
\end{figure}

Against this argument, it is sometimes said (e.g. Strominger, \cite{stro}) 
that we can use
the adiabatic theorem to know everything about the high energy modes: in the
case of expansion of the Universe (where gravity is involved), or for just 
expansion of a box (no gravity involved), we don't need to solve high energy 
physics in order to know what is the present state of modes that are now at 
low energy, but started as superplanckian. We can instead (only for modes 
which have a scale smaller than the scale of change of background geometry
-$H^{-1}$ for cosmological expansion, e.g.) say that if we have an adiabatic
approximation (expansion {\em rate} is infinitely slow), the number of quanta 
is an invariant, or otherwise exponentially suppressed. And then is argued 
that also for black holes, if we would use an "adiabatic time slicing", using
smooth spacelike slices labelled by a time T, (see fig.\ref{fig4})
 everywhere subplanckian, we would 
have  a slow redshift, such as to be able to use the adiabatic approximation.

However, my feeling is that this doesn't say anything, because I want to argue
that particles {\em get created} in the narrow region in fig.\ref{fig4}, 
and afterwards we can use the adiabatic approximation for their {\em evolution}.

The fact is that in any case we will have particles with very high energy in 
the region where they are created. Another point to be noted is that this 
adiabatic time slicing is closely related to observers falling into the horizon
-in some sense, there are "spacelike surfaces of infalling observers", at least 
in some region. But these observers are related by an infinite boost to the 
Schwarzschild observers, so observation in the 2 frames can not be the same
-planckian physics comes in when changing reference frame, so we should have a 
principle of black hole complementarity, as the one advocated in, e.g. 
\cite{lpstu} and \cite{kvv}

\section{The model}

Up to now, I tried to argue for 2 places where planckian physics comes in: the
actual place of {\em creation} of particles, and the jumping from one system
to the other (black hole complementarity) -and that both are related to the 
existence of a horizon-, but now I should argue for what this could bring.

The fact that we have a planckian collision of Hawking particles and ingoing 
particles in the region of the star collapsing through the event horizon was 
argued for a long time by 't Hooft (\cite{th1,th2,th3,th4}), 
however the model he proposes is 
a classical one, based on an Aichelburg-Sexl metric (\cite{as}): 
the shockwave of a massless
particle in Minkowski space boosted at transplanckian momentum p is:
\begin{equation}
ds^{2}=-d\hat{u}(d\hat{v}+4p\log \rho^{2}\delta(\hat{u})d\hat{u})+dx^{2}+dy^{2}
\end{equation}
which can be described (as Dray and 't Hooft showed in \cite{dth}) 
by 2 Minkowski spaces glued with a shift in v, 
\begin{eqnarray}
f=\Delta v&=&-4p\log \rho^{2} (\rho^{2}=x^{2}+y^{2}\; and\; G=\hbar=c=1)\\
&=&16 p(\vec{x}')g(\vec{x}',\vec{x})\\
\nabla^{2}g(\vec{x},\vec{x}')&=&-\delta^{2}(\vec{x}-\vec{x}')
\end{eqnarray}
generalized for the Schwarzschild case as
\begin{eqnarray}
\Delta g -g&=&-2\pi k\delta(\theta), \;k=p2^{9}m^{4}e^{-1}\Rightarrow\\
\delta v&=&kp^{in}(\Omega ')g(\Omega, \Omega')
\end{eqnarray}
Namely, as described by 't Hooft in \cite{th5}, we could consider 
the scattering of
 2 quantum particles with some wavefunction, and describe the interaction as
this classical effect of shifting the Minkowski space, after the shockwave 
passed,
by $\Delta v$. Then we can say that the gravitational interaction is dominant 
at these energies because of infinite time delay of the other types of 
interaction (besides consideration of strength of coupling): $\Delta v
\rightarrow \infty$ for $\rho\rightarrow 0$ ($\rho$ is the impact parameter).

However, it seems to me that this treatment can not be quite true, 
because at these 
ultraplanckian energies we are interested in, at least quantum gravity should
apply, so we should consider at least loop corrections to the classical 
theory.

The picture I want to consider is one where we can say that the high energy 
{\em particles } which are created when the modes leave the star's surface 
interact with the star itself due to the huge relative momentum (or 
equivalently, center of mass energy, $\sqrt{s}$), and so these particles which 
were completely uncorrelated with the particles in the star get correlated by
interaction, so that on the adiabatic time slicing shown before, we can picture
the situation as: the initial state is "$|matter>|Hawking\; radiation>$" 
(quotation marks because Hawking radiation is a density matrix, but we can think
of each state in the density matrix independently since this is not a 
rigorous treatment), which is evolved into 
a state "$\hat{S}|matter>|Hawking\;radiation>$", with some S matrix, 
$<matter\;inside|<modified \;H. rad.|\hat{S}|matter>|Hawking\;rad.>$".

The definition of an S matrix is just a useful approximation, since we don't 
have an initial asymptotic state (the initial  state cannot be said 
to lie at "$-\infty$" and also in some sense Hawking radiation has not been
created yet). But since $|matter\;inside>$ is not accesible outside
before the endpoint of the black hole, the only hope we are left with is the
possible correlations -created by the interaction- between $|modified\;H.rad.>$
and $|matter>$ or at least $|matter\; inside>$. 

In the case of 't Hooft argument, it is quite easy to understand that the only
information we can get is about the momentum of the incoming particles 
(the geodesic shift is proportional to p and we can locally find a reference
frame where the matter has ultraplanckian momentum, not the Hawking radiation).
This is what 't Hooft obtains, but he obtains also an information about the 
electric charge of the matter by considering, besides gravitational interaction,
also electromagnetic interaction, which is also long range, so not affected by 
the infinite shift at small impact parameter.

\begin{figure}
$$\psbox{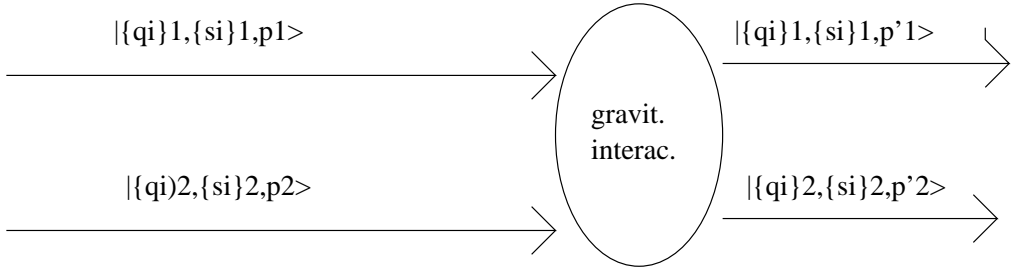}$$
\caption{the case when we have only gravitational interaction at high energies:
charges and spins don't get modified, only momenta do}
\label{fig5}
\end{figure}

Here we can introduce the condition to have the possibility to extract {\em
in principle } the information  about the initial state from the outgoing
radiation. This means that if we consider a scattering matrix as the one 
supposed before, (I emphasize again, in principle we should not be allowed, due
to the initial state, but just as an approximation, for lacking a better tool)
in order to gain information about charges and spins and still have 
gravitational interaction to dominate, we should not only consider quantum 
gravity, i.e. at least loop corrections, but {\em unified quantum field theory 
of gravity and other interactions}. That is because a purely gravitational
interaction will connect only states where all the charges and spins are 
unmodified and only momenta enter the interaction -and get modified. (see fig.
\ref{fig5})
Instead, in a unified quantum field theory of gravity and other interactions, 
we can still have the gravitational interaction to dominate, but the states 
can be in principle different, and moreover, the scattering matrix can 
depend also on these charges and spins, although not in the strength of the
interaction, but instead through polarization tensors. This is what we will
see in the superstring case.

If we obtain such a picture of the interaction in the star, what happens next?
I have tried to emphasize the fact that particles  get created in the star 
and that we have there a significant difference from the Unruh radiation, so 
that particles get created approximately all at once, and independent of the
reference frame, in some sense- that's what we will see also in the adiabatic 
time slicing. That refers though to particle number, and is connected with 
information, because in a region with subplanckian curvature we should be able
to describe the quantum state by particle number.
  Quite a different thing   happens with energy density and energy
flow. 

In \cite{kvv},  a space-time complementarity principle was postulated, saying 
that a permissible semi-classical description of the Hilbert space is possible 
only when the strees-energy fluctuations induce a backreaction on the geometry
smaller than the cutoff used, implying in particular that a measurement inside
the black hole and one outside correspond to noncommuting observables, since 
the matter configurations on which the measurements are made correspond to a
collision in the past history of these configurations which would have had a 
macroscopical backreaction. My opinion is more radical in some sense: they 
{\em have} collided, and the complementarity is related just to the stress 
energy tensor, not to quantum information.  

But let's see first how can we understand what happens to particles (forget 
about energy for the moment). On the adiabatic time slicing, the particles 
of the Hawking radiation and of the matter are seen as separating, but 
there are still correlations between them and with the original particles 
collapsed. They propagate up to the singularity (endpoint), where full
nonperturbative quantum gravity sets in in a way we can't say anything about,
but we may assume (hope?) that there is still a unitary evolution reuniting the
2 sectors.

At this point, let us say again that I have been a bit sloppy in treating the 
interaction, and is actually ment to be just an excuse in lack of a better one 
(nonperturbative, full quantum gravity treatment of the creation of radiation).
First, I considered a scattering matrix, when I should have considered some
nonpertubative treatment, and then put a quantum {\em state}, whereas Hawking
radiation is a thermal state (density matrix). But the goal is just to give an
argument of how to avoid a paradox, not to give a clear solution.

Then, being also sloppy, we can say that the situation can be presented as 
follows (in the adiabatic time slicing)
\begin{eqnarray}
&&|in>=\sum_{i}|in>_{i}\stackrel{Planck\;sca-}{\rightarrow} 
(\; (\sum_{i}|in>_{i})"|Hawking>")\stackrel{le\; int.}{\rightarrow}\\
&&\sum_{j}|in'>_{j}|Hawking'>_{j}\stackrel{Pl.sc.int.at\;endpoint}{\rightarrow}
\sum_{j}|out>_{j}=|out>   
\end{eqnarray}
So that we always have a pure quantum state on the adiabatic slicing, just that 
if we look only on the outside, we see a thermal state, because of tracing over
the inside modes. The correlations we have between the inside and outside states
are just of the type in  the EPR paradox- inherent to quantum mechanics. In
particular, we don't have to use the "no quantum Xerox priciple".

\begin{figure}
$$\psbox{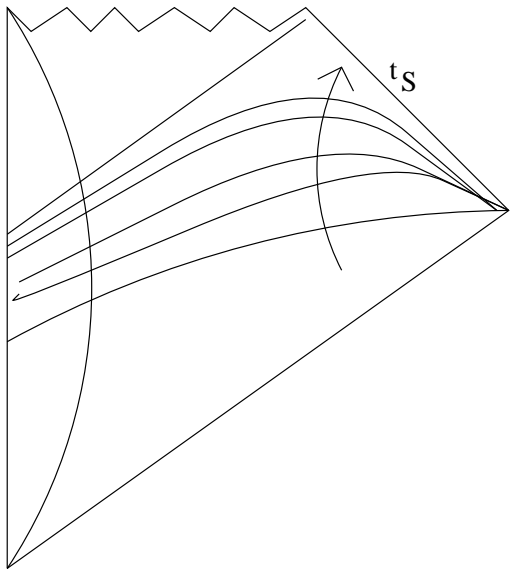}$$
\caption{the time slicing of the Schwarzschild observer}
\label{fig6}
\end{figure}

However, the adiabatic slicing is not the physical one, because outside 
observers see fig.\ref{fig6}, so see more and more of the radiated modes 
(although from the
point of vue of adiabatic time slicing they are there "from the beginning")
and the condition  we put is to have already some information accessible 
concerning the initial state, regarding spins and charges. 

That was about particle number and quantum information. We must also look at
energy density and flow, since we have radiated particles. 

Up to now, all computations of $T_{\mu\nu}$ expectation value where in 2d, and
all seem to rely on the transformation connected with jumping from $|0>_{in}$
to $|0>_{out}$, (\cite{stro,hastro,gid,kaz,tho}) so again 
we have a picture of particle creation at the star's surface:

In 2d models, it is related to the existence of a transformation law between 
$:T_{\mu\nu}:$in one coordinate system and another (related by the schwarzian
derivative) or by using $<T>=\frac{c}{24}R$ and boundary conditions on $I^{-}$
to derive $<T_{\mu\nu}>$ at $I^{+}$ (e.g. Strominger \cite{stro}) or 
because of changing
coordinates between in  and out ones (Birrell and Davies \cite{bd}):

For a metric $ds^{2}=C(u,v)dudv$, the renormalized $T_{\mu\nu}$ is given by:
\begin{eqnarray}
<T_{\mu}^{\nu}[g_{k\lambda}(x)]>_{ren}&=&-\frac{1}{\sqrt{g}}<T_{\mu}^{\nu}[\eta_
{k\lambda}(x)]>_{ren} + \theta_{\mu}^{\nu}-\frac{1}{48\pi}R\delta_{\mu}^{\nu}\;
and\\ \theta_{uu}&=&-\frac{1}{12\pi}\sqrt{C}\partial_{u}^{2}C^{-1/2}\\
\theta_{vv}&=&-\frac{1}{12\pi}\sqrt{C}\partial_{v}^{2}C^{-1/2},\;\theta_{uv}=
\theta_{vu}=0
\end{eqnarray}
which in our case can be rewritten as ($T_{\mu}^{\nu}[\eta_{k\lambda}(x)]=0$):
\begin{eqnarray}
<0|T_{uu}|0>&=&<0|T_{vv}|0>=-F_{u}(C)=\frac{1}{192\pi}[2CC"-C'\;^{2}]\label
{flux1}\\<0|T_{uv}|0>&=&\frac{1}{96\pi}CC"\;where\label{flux2}\\
F_{x}(y)&=&\frac{1}{12\pi}y^{1/2}\frac{\partial^{2}}{\partial x^{2}}(y^{-1/2})
\end{eqnarray}
So we have a vacuum polarization due to the curved space, and when matching the
in coordinates and out coordinates:
\begin{eqnarray}
\hat{u}&=&\beta[\alpha(u)-2R_{0}], \hat{v}=v \Rightarrow\\
ds^{2}&=&C(\hat{u},\hat{v})d\hat{u}d\hat{v},\;\hat{C}(\hat{u},\hat{v})=C(r)
\frac{du}{d\hat{u}}\frac{dv}{d\hat{v}}
\end{eqnarray}
so that the expectation value of the out energy-momentum tensor in the in
vacuum gets an extra factor:
\begin{eqnarray}
<\hat{0}|T_{uu}|\hat{0}>&=&previous+(\alpha')^{2}F_{U}(\beta')+F_{u}(\alpha')\\
<\hat{0}|T_{vv}|\hat{0}>&=&and \;<\hat{0}|T_{uv}|\hat{0}>=previous
\end{eqnarray}
This extra term, giving the Hawking radiation, tends to $\frac{k^{2}}{48\pi}$
when $R(\tau)\rightarrow R_{h}$ (C$\rightarrow$0, or $u\rightarrow\infty$),
i.e. at late times. 

Then the energy density and energy flux are well defined outside the black hole,
however, {\em on the horizon}, the correct procedure is to analyze what a 
fiducial observer sees, and he measures an energy density $<T_{\mu\nu}>u^{\mu}
u^{\nu}$, where the velocity of a freely falling observer of fixed spacial 
Kruskal 
coordinate $\bar{r}$ is $u^{\mu}=(\frac{1}{\sqrt{\bar{C}}},0)$ in $\bar{r},\bar
{t}$ coordinates ande the metric in Kruskal coordinates $\bar{u},\bar{v}$ is
$ds^{2}=\bar{C}d\bar{u}d\bar{v}$. So that he measures
\begin{eqnarray}
\frac{1}{\bar{C}}<\hat{0}|T_{\bar{t}\bar{t}}|\hat{0}>&=&<\hat{0}|T_{\bar{u}}
^{\bar{v}}+T_{\bar{v}}^{\bar{u}}+2T_{\bar{u}}^{\bar{u}}|\hat{0}>\;and\\
T_{\bar{u}}^{\bar{v}}&=&T_{u}^{v}e^{t/2M}=T_{uu}\frac{e^{t/2M}}{1-\frac{2M}{r}}
\\ T_{\bar{v}}^{\bar{u}}&=&T_{v}^{u}e^{t/2M}=T_{vv}\frac{e^{-t/2M}}{1-\frac{2M}{
r}}\\
T_{\bar{u}}^{\bar{u}}&=&T_{u}^{u}=T_{uv}\frac{1}{1-\frac{2M}{r}}
\end{eqnarray}
and by putting $C=(1-\frac{2M}{r})$ in equations \ref{flux1},\ref{flux2},we get:
\begin{eqnarray}
<\hat{0}|T_{\bar{u}}^{\bar{u}}|\hat{0}>&=&-\frac{M}{24\pi r^{3}},\\
<\hat{0}|T_{\bar{v}}^{\bar{u}}|\hat{0}>&=&\frac{1}{48\pi e^{t/2M}(\frac{r}{2M}
-1)}[\frac{1}{r^{2}}-\frac{3}{2}\frac{M}{r^{3}}]=\\&&
\frac{M}{3\pi}[\frac{1}{r^{3}}-\frac{3}{2}\frac{M}{r^{2}}]\frac{e^{r/2M}}{
\bar{v}^{2}}
\end{eqnarray}
which are both finite at the horizon (r=2M, $\bar{v}$=finite, $t\rightarrow
\infty$), however, in $<\hat{0}|T_{\bar{u}}^{\bar{v}}|\hat{0}>$, both the 
term coming from vacuum polarization,
\begin{equation}
<\hat{0}|T_{\bar{u}}^{\bar{v}}|\hat{0}>|_{vac}=\frac{e^{t/2M}}{1-\frac{2M}{r}}
[\frac{M^{2}}{16\pi r^{4}}-\frac{M}{24\pi r^{3}}]
\end{equation}
and the one coming from the particle,
\begin{equation}
<\hat{0}|T_{\bar{u}}^{\bar{v}}|\hat{0}>|_{part}=\frac{e^{t/2M}}{1-\frac{2M}{r}}
\frac{k^{2}}{48\pi}
\end{equation}
are infinite at the horizon, however, the one coming from the vacuum 
polarization is negative and cancels the infinity, giving a finite result:
\begin{equation}
<\hat{0}|T_{\bar{u}}^{\bar{v}}|\hat{0}>=\frac{1}{48\pi 16M^{2}}\frac{\bar{v}^
{2}e^{-r/2M}}{8Mr}(1+\frac{4M}{r}+12\frac{M^{2}}{r^{2}})
\end{equation}

and the freely falling observer measures a finite energy density, because as we 
see, the vacuum polarization cancels the infinity arising due to particles.

So we see that the picture is of a black hole complementarity principle, due to 
the infinite boost and relative acceleration of the fiducial and freely falling
observers at horizon:

The Schwarzschild observer sees particles going away and carrying energy, but
for him the interior of the black hole is blocked by an infinite boost. He sees
information flow and energy flow, and the 2 have the same source.

The freely falling observer sees energy density and energy flow, but 
gravitational interaction hides the infinity of the energy density. A consistent
treatment of this frame jump would require also full quantum gravity. For
this observer, the outgoing particles are spread over a large distance -in the
adiabatic time slicing- (because the outgoing modes at infinity have an average
size equal to the size of the black hole, in the adiabatic slicing we will 
also have modes spread over the whole slice) and on this adiabatic slice there 
is a correlation of the outgoing modes with the incoming and the original ones.

A complementary view was taken in \cite{lsu,lpstu}, 
where on this "nice slice" (adiabatic
slice) was proven that, although local fields seem not correlated, superstring 
fields are correlated, because of the enhancement of the commutator of the
fields by an infinite boost. So there, the correlation was the result of the
infinite boost connected with black hole complementarity, however, we could also
trace the moment where the correlation {\em begins} with the collapsing 
through the event horizon.

\section{The superstring calculation}

The case relevant for us is the scattering of 2 massless particles having 
10d polarization tensors $\epsilon_{a}, \epsilon_{b}$. As I emphasized, this
may not be exactly applicable in the case we are studying, since we can't
define rigorously an S matrix (there are no asymptotics in the in region),
however this is the best tool at hand. Moreover, since in the regions of 
interest the curvature is subplanckian, we could try a flat space computation
(Minkowski space), and see what happens (we can say that the "Rindler 
approximation" provides us with a reference frame).

First of all, because of supersymmetry in the higher dimensional theory,
we expect that the exact amplitude will contain the polarization tensors in 
the kinematic factor, which e.g. for closed strings and external gravitons
is (for other superstring theories it should be the same type of situation):
\begin{eqnarray}
K&=&\epsilon_{a}^{\mu_{1}\mu_{2}}\epsilon_{b}^{\nu_{1}\nu_{2}}\epsilon_{c}^{\rho
_{1}\rho_{2}}\epsilon_{d}^{\lambda_{1}\lambda_{2}}K_{\mu_{1}\nu_{1}\rho_{1}
\lambda_{1}}K_{\mu_{2}\nu_{2}\rho_{2}\lambda_{2}}\\
K_{\mu\nu\rho\lambda}&=&-\frac{1}{4}(st\delta_{\mu\rho}\delta_{\nu\lambda}+
su\delta_{\nu\rho}\delta_{\mu\lambda}+tu\delta_{\mu\nu}\delta_{\rho\lambda})
\\&+&\frac{1}{2}s(k_{1\mu}k_{2\rho}\delta_{\nu\lambda}+k_{3\nu}k_{1\lambda}
\delta_{\mu\rho}+k_{3\mu}k_{2\lambda}\delta_{\nu\rho}+k_{4\nu}k_{1\rho}\delta_
{\mu\lambda})\\&+&\frac{1}{2}t(k_{1\nu}k_{3\lambda}\delta_{\mu\rho}+k_{4\rho}
k_{2\mu}\delta{\nu\lambda}+k_{4\nu}k_{3\mu}\delta_{\rho\lambda}+k_{1\rho}
k_{2\lambda}\delta_{\mu\nu})\\
&+&\frac{1}{2}u(k_{2\mu}k_{3\lambda}\delta_{\nu\rho}+k_{4\rho}k_{1\nu}\delta_{
\mu\lambda}+k_{4\mu}k_{3\nu}\delta_{\rho\lambda}+k_{2\rho}k_{1\lambda}\delta_{
\mu\nu})
\end{eqnarray}
The (present form of?) superstring theory has the particularity that all the 
known particles appear as massless from 10d point of view, so that they 
correspond to the vacuum, the only way of discerning between them being via
polarization tensors. So in particular, charges and spins are expected to be
described also by polarization tensors, in some way. That means that the 
superstring amplitude for scattering of some 4d fields will depend upon the 
4d charges not via the coupling, but via polarization tensors. 

\begin{figure}
$$\psbox{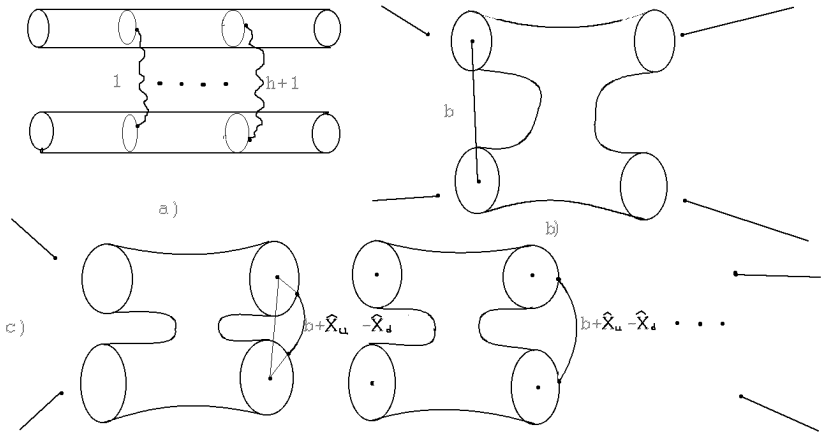}$$
\caption{a)in the h loop diagram, the main contribution comes from exchange 
of gravitons; b)the tree amplitude can be rewritten as a scattering at 
displaced  impact parameters; c)the loop diagrams are interpreted as 
rescatterings at displaced impact parameters, which is nontrivial}
\label{fig27}
\end{figure}

In order to make a comparison with the classical picture of 't Hooft, we look
at a domain where to be able to do perturbation theory. We want s to be above
the Planck scale, but to be still in the perturbative domain involving loops.
This means that the loop expansion parameter in D dimensions, $\alpha_{D}=
\alpha_{10}\frac{\lambda_{S}^{10-D}}{V}$ ($\alpha_{10}=k^{2}$, V=volume of 
compactified coordinates and $\lambda_{S}$= string length) 
should be $\alpha_{D}\ll 1$, but $E_{CM}>E_{Pl}$
so $s>\lambda_{Pl}^{2}$, and $\lambda_{Pl}^{2}=\alpha_{D}\frac{\lambda
_{S}^{2}}{32\pi}$ implies $\alpha_{D}(\alpha's)>1$, so that we must have 
$\alpha's\gg 1$. Also, the momentum transfer should be small and fixed,
$|t|<1/\alpha'$. Then, since 
\begin{equation}
A(s,t)\sim\sum_{poles}\frac{s^{\alpha(t)}}{t-M_{t}^{2}}
\end{equation}
the leading behaviour is given by the exchange of the particle with the largest
$\alpha(t)$, i.e. the graviton at $\alpha(t)=2+\alpha'/2t$. So in this regime
we can again say that the gravitational interaction dominates (actually,
dominates anyway). The full analysis of this case was made in 
\cite{acv1,acv2,am}, however
there is an intuitive -and nonrigorous- way of deriving the result. If we
define an impact parameter transform, extracting out the polarization tensors
(the interesting part, after all) e.g., in the limit $s\rightarrow \infty$,
t fixed, when K becomes $K=(\frac{s}{2})^{4}\epsilon_{a}* \epsilon_{d} 
\epsilon_{b}*\epsilon_{c}$, define:
\begin{equation}
\frac{1}{s}A(s,t)=\epsilon_{a}*\epsilon_{d} \epsilon_{b}*\epsilon_{c} 4\int
d^{D-2}\vec{b}e^{i\vec{q}\vec{b}}a(s,b)
\end{equation}
then we can rewrite for a(s,b):
\begin{equation}
a_{tree}(s,b)=<0|\int\frac{d\sigma_{u}d\sigma_{d}}{(2\pi)^{2}}:a_{tree}(s, \vec{
b}+\hat{\vec{X}}^{u}(\sigma_{u})-\hat{\vec{X}}^{d}(\sigma_{d})):|0>=<0|\hat
{\delta}|0>
\end{equation}
because the VEV invoves only $<0|:e^{i\vec{q}(\vec{X}^{u}(\sigma_{u})-\vec{X}
^{d}(\sigma _{d}))}:|0>$, which is 1 due to normal ordering ( and because we 
define $\hat{\vec{X}}^{u},\hat{\vec{X}}^{d}$ not ot contain 0 modes). The 
physical interpretation is that we can understand the scattering as taking
place at displaced impact paratmeter and the VEV means that the external states
are vacuum states. (see fig.\ref{fig27} b)

\begin{figure}
$$\psbox{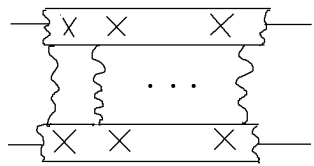}$$
\caption{rescattering: the intermediate states are on-shell}
\label{fig30}
\end{figure}

When we consider the diagram with h loops, the main contribution comes from the
exchange of h+1 gravitons: (see fig.\ref{fig27}a), which can also be 
reinterpreted as a row of rescatterings, as in fig.\ref{fig27}c- so that the 
internal propagators are on shell (the crosses put on the diagrams in 
fig.\ref{fig30}), each of which is a tree amplitude, but on the internal
propagator we have {\em any} kind of states, so that we don't have products
of trees, 
\begin{equation}
a^{(h)}(s,b)\neq <0|\hat{\delta}|0><0|\hat{\delta}|0>...=(<0|\hat{\delta}|0>)
^{h+1}
\end{equation}
but instead we have:
\begin{equation}
a^{(h)}(s,b)=<0|(\hat{\delta} )^{h+1}|0>=<0|\hat{\delta}\sum_{i}|i><i|\hat{
\delta}...\sum_{i}|i><i|\hat{\delta}|0>
\end{equation}
When we connect the tree amplitudes however, permutations of the $\hat{\delta}$
's are redundant, so we  should have an $\frac{1}{(h+1)!}$, and there are 2 
ways of connecting 2 trees, so we have an $2^{n}$, and the 1 loop amplitude 
is purely imaginary in this limit, so we have an i for each loop, so:
\begin{equation}
a^{(h)}(s,b)=\frac{(2i)^{h}}{(h+1)!}<0|(\hat{\delta})^{h+1}|0>
\end{equation}
If we resum the leading contribution of all diagrams, we get:
\begin{equation}
a(s,b)=\sum_{h}a^{(h)}(s,b)=<0|\frac{1}{2i}(\hat{S}(s,b,\hat{\vec{X}}^{u},
\hat{\vec{X}})-1)|0>
\end{equation}
where $\hat{S}=e^{i\delta}$can be interpreted as an S matrix. Amati and Klimcik
(\cite{ak}) found also that if you solve the string equations of motion in a 
shockwave background of a type similar to Aichelburg-Sexl:
\begin{equation}
ds^{2}=-2dudv+f(x)\delta(u)du^{2}+dx^{2}
\end{equation}
and define a scattering matrix relating free in (before the shockwave) and 
out (after the shockwave) modes by $S^{+}a_{in}S=a_{out}$ (a=$p^{u},x^{i},v,
\alpha_{n}^{i},\tilde{\alpha}_{n}^{i}$), then:
\begin{equation}
\hat{S}=exp(\frac{i}{2\pi}p^{u}\int_{0}^{\pi}d\sigma\;f(\hat{X}(\sigma^{u})))
\end{equation}
If we formally put 
\begin{equation}
f(y)=-q^{v}\int_{0}^{\pi}\frac{8}{s}:a_{tree}(s,y-\hat{X}^{d}(\sigma^{d},0):
\frac{d\sigma_{d}}{\pi}
\end{equation}
(not very well defined-I have an operatorial expression in a scalar function,
the shockwave profile), we get formally the same expression as for the 
interacting superstring case, so that we can think of the interaction as 
producing some sort of shockwave effective metric with respect to one of the
strings. And if we neglect X, (higher string corrections), and look for 
$y\gg\sqrt{\alpha'\log s}$, we get (Amati and Klimcik \cite{ak})
\begin{eqnarray}
f(y)&=&-q^{v}\frac{16\pi G_{N}}{(D-4)\Omega_{D-2}}\frac{1}{|y|^{D-4}}\\
\Omega_{d}&=&\frac{2\pi^{d/2}}{\Gamma(d/2)},\;and\\
f(y)&=&-q^{v}8\log |y|\;in\;D=4
\end{eqnarray}
which is Aichelburg-Sexl metric.

So we can see that after we factor out the kinematical factor -the interesting 
part- what is left is something which looks like the Aichelburg-Sexl metric
at large impact parameters, but if we look at smaller impact parameter we see
an extra bonus: the presence of $\hat{X}$ means that we will have a dependence
on the external states- for external particles which are light compared to s,
(so as to still have our analysis to be applicable), but still massive, the 
result will be in general different. This doesn't help our case because for the 
black hole the heavy states, even if they are generated, cannot leave the black
hole (fall back again), so they are not seen "at infinity". Still, this hints
that a consistent computation  will have the scattering matrix depend upon all
the information in the initial state (as it should in a unified theory of 
everything), and moreover, the gravitational interaction could still dominate
and the long distance behaviour of the interaction  be given by the Aichelburg-
Sexl type of metric.

As we saw, in the limit $s\rightarrow\infty$, t small and fixed, 
$K=(\frac{s}{2})^{4}\epsilon_{a}*\epsilon_{d}\epsilon_{b}*\epsilon_{c}$, so 
we are in some sense again where we started, because this depends only on the 
coupling of the initial incoming particle (a) to the modified particle (d)
and of the initial Hawking particle (b) to the modified one (c), and this would 
have been expected anyway. But this case was only ment to be a tool to look
at small perturbations (small t). If instead, we look at big t, the kinematic
factor will now contain all sorts of couplings, so when we sum over the initial
Hawking particles (b) (as nonphysical-they don't appear in reality), the 
scattering amplitude of the modified Hawking particle will depend on the
incoming particle and the modified one (behind the horizon).

That means that some information on the initial state will be already available
in the spectrum and correlations of Hawking radiation, and this will also be
related to the states behind the horizon.

The planckian endpoint (where full, nonperturbative quantum gravity applies),
can bring then whatever (a remnant, radiation of the inside states, or just 
dissapearence of them). The states inside the horizon don't matter, in some 
sense, because it is the EPR-type of paradox which appears, so if the state
outside is known, the state inside is known also.

The information on the initial state can be found in the outgoing radiation,
but just that the "strength" of the dependence on the initial state increases
with time (with higher advanced time, the blueshift of the outgoing radiation
when being in the interacting region is growing, so the effects should also 
grow). The condition to have information available is equivalent to having a 
unified quantum field theory of gravity and other interactions at high energies.

\section{Conclusions}
I have tried to argue that the black hole information paradox can be avoided,
because quantum gravity appears, besides the planckian endpoint, in 2 different
places: in the black hole complementarity principle, fiducial  and freely
falling observers having relative boosts going quickly beyond Planck scale
(tend to infinity), so the 2 pictures need not be the same, and in the moment 
of creation of particles, where there is a superplanckian interaction with the 
matter. In order to have all information available outside the horizon, about 
the infalling state, we should have a unified quantum field theory of gravity
and other interactions at high energies. Superstring theory passes the test, 
and in some limit, we find the classical picture emerging.
\section{Acknowledgements}
I would like to thank the Niels Bohr Institute for supporting my stay here,
making possible the completion of this work, and I would like to thank also
 my supervisor, Prof. Poul Olesen, for useful discussions and comments.

\end{document}